# Novel exact solutions of the Duffing equation: stability analysis and application to real non-linear deformation tests

A.D. Berezner [a*], V.A. Fedorov [a], N.S. Perov [b], G.V. Grigoriev [a]

[a] Theoretical and Experimental Physics Department, Derzhavin Tambov State University, Tambov 392000, Russia

[b] Magnetism Department, Lomonosov Moscow State University, Moscow 119991, Russia

[*]a.berezner1009@gmail.com

**Abstract**

In this study, novel exact solutions of the Duffing equation with their phase portraits have been proposed and reasoned. It is shown that phase trajectories are initially elliptical and become distorted in the unstable area within the growth of the variable parameter. Instability criteria of identified solutions have been determined together with the Fourier series transformation up to the first and high harmonics in a sense of the physical interpretation. An explicit form for the differential operator, corresponding to considered functions, has been derived with evaluation of its main functional spectrum. Non-isothermal creep tests of different materials were completely described using the Duffing equation via noted solutions up to the fracture as processes with personal deformation response. We successfully examined a relationship between the thermal and magnetic properties of the ferromagnetic amorphous alloy under its non-linear deformation, using the critical exponents. With a high linear correlation between our model and experiments, behaviour of organic and metallic systems is well predicted at the same thermo-mechanical testing conditions on the mesoscale.

**Keywords:** non-linear processes, Duffing equation, creep, viscoelasticity, critical exponents, phase transitions

# 1. Introduction

A quantitative description of various non-linear processes using basic mathematical functions (or their superposition) remains incomplete in the fields of applied mathematics, physics, and technics [1-8]. In this case, the presence of numerous key accompanying factors [9,10] and their complicated relations [11,12] becomes the main problem to obtain necessary information. Furthermore, for some differential equations, describing such processes, there are no main methods to obtain an explicit solution, and one can apply complicated special functions [13] that affect practical versatility and quantitative model prediction [14]. In some cases, an exact solution is expressed only with local special functions [15] whose application cannot be practically effective, and approximation becomes more reasonable [16]. The Duffing equation [17,18], which is a homogeneous (or inhomogeneous) ordinary second-order non-linear differential one, describing damped (undamped) oscillations, can be mentioned for example. A physical or engineering illustration for this equation is an ideal dynamical system consisting of a parallel-connected inelastic spring and a non-linear viscous cylinder pump [19]. An effective interpolation or analytical derivation of a solution for the differential equation is the direct method to describe oscillator behaviour. Clear understanding of transition conditions between linear and nonlinear stages becomes possible with explicit and mathematically simple solutions [20]. From theoretical and applied perspectives, analysis of the phase portraits and stability at any given solution [21] with variation in its free parameters is supposed to be also important.

Phase transitions (or critical phenomena) can cause the noted nonlinearity, for example in polymers and alloys [22,23]. In experiments, such effects appear at differential scanning calorimetry [24] and in different diffraction patterns [26]. Some effects are described in phase diagrams (component mass vs. temperature and so on) [27]. In numerous systems, heating rate and other driving parameters, for instance, describing with amplitude-frequency response (AFR) in mechanical testing, make an additional impact on nonlinear processes [28]. Despite the proposed well-known theory of second-order phase transitions [29,30], its using can be ineffective in certain cases (also for the first-order transformations) due to a personal system complexity (like renormalization groups, etc.). Therefore, a proposal of new model approaches, according to critical exponents at nonlinear conditions, seems to be necessary. Herewith, a combination of several theoretical concepts (mechanical with thermodynamic, for instance) is complementary, and it could become a basement for the comprehensive model description [31]. Also known that different testing machines, used for the dynamo-mechanical (DMA) and thermo-mechanical (TMA) analysis, impose the upper limit of the heating rate [32] that must be considered in real nonlinear experiments, involving ten times faster temperature reach and multiple phase transitions.

As we mentioned above, deformation of materials involves both viscoelastic [33] and nonlinear [34] scenarios. For instance, non-isothermal creep [35,44] and DMA [37] are nonlinear deformation experiments due to the absence of the exponential elastic response in their plots, and the monotone heating combined with oscillating mechanical load causes resonance effects with attractors in phase portraits [38]. Apart from proposed relationships for thermo-mechanical conditions [39,40], a suggested main differential equation with stability analysis of solutions is necessary for non-isothermal deformation. Universal modelling of some critical effects, arising in different systems under the same thermomechanical conditions, is also interesting with respect to possible switching between the special cases during the whole process.

Thus, we set the goal here to determine the main nonlinear differential equation describing thermo-mechanical experiments with constant or oscillating load and different heating modes. Analysis of this equation and its solutions with respect to mathematical stability at linear or nonlinear deformation tests is the basic task for this study. Fractography of materials, undergoing the creep at a constant heating rate with phase transition, will be separately considered.

## 2. Experimental procedure

Aside from the model analysis, conducted in frames of the main nonlinear differential equation and its solutions, probative empirical data were collected with laboratory equipment to study non-isothermal creep [35]. We used amorphous magnetic alloys (also metallic glasses or MG, at.wt.%, AMAG-183: Co-82.69%, Fe-2.21%, Si-7.77%, Mn-4.19%, B-2%, Cr-1.14% and AMAG-200: Fe-80.22%, Si-8.25%, Cu-1.44%, Nb-10.09%), manufactured like Vacuumschmelze or Metglass products, and polycrystalline ones (pt. wt.%: Al-98%, Fe-2%) together with non-crystalline magnetic micro-rods (pt. wt.%: Co-75%, Ni-0.9%, Fe-6%, Si-8%, B-7%, Cr-31%) as testing specimens. To compare with alloys, the deformation behaviour of an organic structure (a silk fibre) was estimated under the same temperature conditions. The size of AMAG-183 and AMAG-200 ribbons was about 55×3.5×0.015 mm by length, wide, and thickness, respectively, while polycrystalline one was made in a 50×6×0.01 mm configuration. Magnetic micro-rods (or just named «rods» here) of 3 μm in diameter by themselves were initially fabricated inside a cylinder glass cover about an 8 μm radius (see Fig.1a) with the possible change up to 50 mm in length.

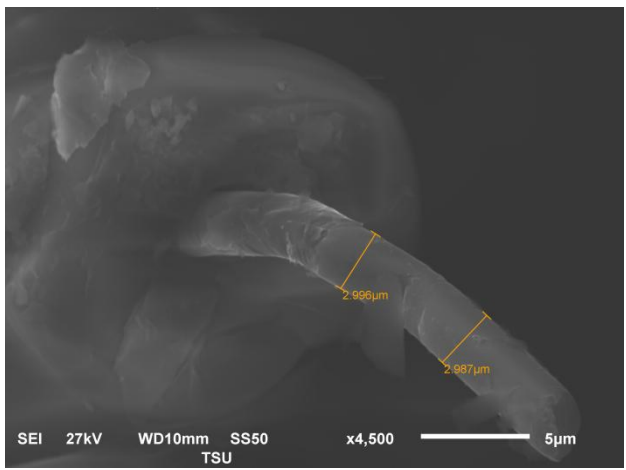 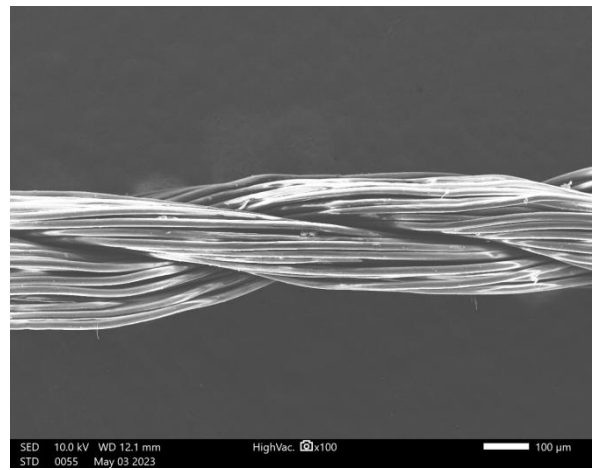

a)                                          b)

Fig.1. Geometry and structure of the: a) micro-rod; b) silk fibre

The silk fibre has a wicker rope-like structure with the maximal diameter of about 267 μm and 16 μm in thickness for a one filament (see Fig. 1b).

During non-isothermal creep tests of AMAG alloys and silk fibres, the $V_T=Const$ fixed heating rate was reached as 1, 2 or 3 K/s with applied 1, 17 or 42 N constant loads for MGs but 0.1 N for the organic system. Crystalline alloys were tested using the same $V_T$ values, but with applied loads of 1, 1.5 or 3 N. The amorphous magnetic rod manifested the similar creep scenario at $V_T =1$ K/s and 0.1 N of load. The mentioned difference between loads and heating rates is caused by the tensile stress of materials as a limit for the possible experimental conditions. For all the mentioned investigations, initial factors (i.e. the heating rate and load) were selected either separately or together (except the rods with values chosen once) to determine the system stability against the external impact. In summary, all the mentioned tests together were carried out about 1000 times for a better understanding of the statistical appearance in behaviour.

To structural testing of specimens before and after experiments, we used D2 Phaser and XRD-7000S X-Ray diffractometers. The melting and polymerisation temperatures of a silk fibre (with 1.426 mg mass) were determined using an EXSTAR TG/DTA 7200 complex both in the air atmosphere and a carbon dioxide stream (at the 50 ml/min flow rate) on a platinum substrate. The amorphous ribbons were annealed with a SNOL 8.2/1100 furnace for 10, 25, and 38 minutes. Fractography, plastic flow, and melting were examined using optical or scanning electron microscopy (Vega3 Tescan and JCM-7000 Jeol). Magnetic properties of metallic glasses were studied with a LakeShore-7407 magnetometer under ±125–1000 Oe external field both in parallel and perpendicular projections relative to specimens.

## 3. Results and discussion

*3.1. Model basement and analysis*

To enable a universal analysis of various non-linear thermo-mechanical processes (like non-isothermal creep here or DMA in [28,38,39]), we propose some exact solutions for the master differential equation. Therefore, it is necessary to formulate and prove two basic theorems on explicit solutions.

Let us define the $x(t) = x_0 + \dfrac{Ct}{B^2 - Bt}$ (1) function with positive $x_o$ constant, $C$ and $B$ parameters, and $t$ variable, determined in a $[0;B)$ half-open interval. Thus, there is a following theorem.

<u>*T.1.*</u> Function (1) is an exact solution of the $\ddot{x} - P_3(x) = 0$ (2) non-linear homogeneous differential equation, where $P_3(x)$ is a cubic polynomial by $x$ variable with $-6/B^2$, $-6/CB$, $-2/C^2$ constant coefficients at $x$, $x^2$, and $x^3$ positive terms, respectively, with $-2C/B$ one as the free term.

The proof for this theorem is provided in the **Appendix** (A.1) section. Similarly to <u>*T.1.*</u>, formulation of the second theorem about an explicit solution is possible.

Let the $x(t) = x_0 + \dfrac{Ct}{B^2 - Bt} - b\cos(\omega t)$ (3) function be defined in the $[0;B)$ half-open interval, where the first two terms form (1) with $b$ and $\omega$ as positive constants.

<u>*T.2.*</u> Function (3) is an exact solution of the $\ddot{x} - P_3(x;t) = \gamma \cos(\omega t)$ (4) non-linear inhomogeneous differential equation, where $P_3(x;t)$ is a cubic polynomial by $x$ and $t$ variables with $-6/B^2$, $-6/CB$, $-2/C^2$ constant coefficients at $x$, $x^2$, and $x^3$ positive terms, respectively, with $-2C/B$ one as the free term, and $\gamma = \omega^2 b$.

Reasoning for <u>*T.2.*</u> is similar to the proof for <u>*T.1.*</u>, considering the transfer of harmonic term (from the right side in (4) to the left one), we get the final identity 0=0 (see (A.2) in **Appendix**).

The physical and engineering meaning of (2) is an undamped Duffing non-linear differential equation (i.e. $\ddot{x} + c_1 x + c_2 x^2 + c_3 x^3 = 0$, where $c_1 - c_3$ are constants), describing free oscillations of an inelastic spring parallel combined with a piston at its periodical viscous rubbing in a cylinder (in case (4), we can notice a damped pendulum $\ddot{x} + c_1 x + c_2 x^2 + c_3 x^3 = \gamma \cos(\omega t)$) [19]. Accurate to the $m$ mass, the meaning of (1) can be clearly shown, i.e. within the growth of heating temperature, an elastic reaction of the specimen weakens and deformation accelerates to the third-order value by $x$ in $t$ time. The stability point shifts rapidly toward fracture, and viscosity of the material can generate beating, resonance, and other phase effects (bifurcations) [39]. As a result, in experiments [28,38,39], the rational function (1) is accurately applied across the entire mesoscale. A similar conclusion is valid for the equation (4) with sine or cosine in the right side. Herewith, there are several ways to represent this equation. In original Duffing work [17], the term involving the first time derivative of coordinate is not mentioned, but it can be generally included in representations [18,19]. By using (1), the $x$ or $x^2$ term is replaced with a non-linear expression of $\dot{x}$. To achieve this, we may use the substitution in relative form $\varepsilon = \sqrt{\dfrac{C\dot{\varepsilon}}{x_0} - \dfrac{C}{Bx_0}}$ (5) [35], wherein $\varepsilon = \Delta x / x_0 = (x - x_0)/x_0$.

Already known solutions such as Jacobi and Weierstrass elliptic functions [14-16] describe the viscoelastic region, and (1) or (3) provide modelling for elastic and plastic deformation flow with the further fracture of specimens ($x_o$, $B$, $C$, $\omega$, $b$ are initial length, fracture time, flow coefficient, cyclic frequency, and loading amplitude, respectively). A replacement of the $x^3$ term using (5) yields an expression for non-linear (non-Newtonian) viscosity as $\dot{x}$ appears with a 3/2 power there, while $x^2$ describes linear internal friction. Also (1) and (3) can be transformed to the viscoelastic deformation function (Maxwell-Voigt model [33]) in the limit transition. To achieve this, we need the reaction force function (Eq. (7) from [39]):

$$F_{reaction} = F_{load} - \gamma \sin\left(\omega \frac{B^2 \Delta x}{C + B\Delta x}\right) - \frac{2mC}{\left(B - \frac{B^2 \Delta x}{C + B\Delta x}\right)^3}, \text{ derived using (1) and (3) in}$$

frames of the second Newton's law, whose special case is the Duffing equation, i.e. the $-P_3(x) - \gamma\cos(\omega t)$ term describes the specimen response. Then, $F_{reaction} \to 0$ at $\Delta x \to 0$, and, in order of smallness, the term $-k\Delta x = -6m\Delta x/B^2$, being material elastic force, will slower tend to zero in comparison with others, i.e. $F_{reaction} \approx F_{elastic}$ (also, some analysis on this can be found in [39]). Therefore, with respect to the mentioned reasoning, the master Duffing equation becomes to the $\ddot{x} - kx \approx 0$ form, whose solution is known as Maxwell-Voigt viscoelastic function. Thus, the reaction force function transforms to elastic one while non-linear (1) or (3) solutions tend to $x(t) = x_0(1 - \exp(-qt))$ isothermal functional case with $x_0$ and $q$ parameters. Note, with temperature decrease comparable to $V_T$, curves (1) and (3) demonstrate relaxation like isothermal ones, and that testifies about relation between viscoelasticity and proposed solutions. Finally, the Duffing equation (2) becomes to Maxwell or Kelvin-Voigt models with equality to zero of terms, containing higher $x$ coordinate powers (>1), and $q \sim \frac{E}{\eta\dot{\varepsilon}} \sim \frac{E}{\eta\dot{\varepsilon}(V_T)} \to 0$ if $V_T \to \infty$, where $E$, $\eta$ are Young's modulus and dynamic viscosity, respectively, also $\dot{\varepsilon}(V_T)$ is the derivative of (1) or (3) in $\varepsilon = \Delta x / x_0$ relative form with the further substitution $t(V_T)$. For instance, in discussed experiments, $t(V_T)$ can be explicitly determined as a linear [35] or another elementary function that allows for the limit transition.

Besides (1) and (3) solutions, their separate superposition with a linear function $\pm vt$ is also acceptable as the second-order derivative by time completely eliminates the linear term, and a proof for applicability becomes similar to the described one in (A.1). Thus, in this study, three major types of solutions for the Duffing equation are proposed with the further discussion of their application in different cases. From Eq. (5), one can obtain the analytical form of a differential operator $A$, i.e. $Ax = dx/dt = (x + M)^2 \cdot N$, which differs from $Ax = dx/dt = \lambda x$ one in linear algebra (where $\lambda$ is an eigenvalue of $A$, $M=C/Bx_o$, $N=x_o/C$). This non-linearity of $A$ permits estimating its basis functions, which generate solutions to (2) or (4), and whose expression has arithmetic, trigonometric, and hyperbolic superpositions by $t$ [41], partially related to (1) or (3), as well as with some Jacobi and Weierstrass elliptic functions. But an available representation of the basic set is only supplementary (or qualitative) as its functions themselves do not satisfy equations (2) or (4) unlike their superpositions. Moreover, (1) and (3) are related to a differential form defined in the $x$-$t$ plane [40] with some peculiarities [28], observed during its derivation. And functions, distinguishing from (1) by added terms (i.e. continuous differentiable superpositions of $f(t)$ in the $[0;B)$ half-open interval), form a subset of solutions like (4) with the similar proof (like (A.2) in **Appendix**). For example, equation:

$$f_1(t) = x(t) + f(t) = x_0 + \frac{Ct}{B(B-t)} + x_0(1-\exp(-qt)) + x_0 \sqrt[\beta]{\ln\left(\frac{B}{B-t}\right)} - b\cos(\omega t) - vt, \quad (5a)$$

where the fourth term corresponds to the Weibull distribution [42] with the $\beta$-parameter ($v$ represents the shrinkage velocity), and the whole function (5a) is a solution to $\ddot{x} - P_3(x;t) = M(\cos(\omega t); -\exp(-qt); \ln(B/(B-t)))$ whose partial cases describe isothermal or non-isothermal creep, DMA, TMA as well as fatigue experiments. As we mentioned above, the main condition $V_T \to \infty$ for a non-isothermal experiment change the third term in (5a) to $x_o$, and the $\sqrt[\beta]{\ln\left(\frac{B}{B-t}\right)}$ function tends to zero with the fixed $\beta$ in a finite temperature interval $\Delta T$. For isothermal experiments, (5a) is surely considered without a $t=f(V_T)$ relationship, when constant $C$, $B$, $q$, $\beta$, $\omega$ parameters will determine the sigmoid deformation curve.

During analysis of differential equations, their stability (by Lyapunov or Poincare), phase portraits (for cyclic solutions), and resonance conditions (perturbations, beating, etc.) [19,21] can be considered, and that assists in a universal description of different processes. For (3), analytical set of phase portraits in $x, \dot{x}$ coordinates are supposed to be appropriate. Particularly, $cos(\omega t)$ and $sin(\omega t)$, in (3) with derivative, respectively, after a separate powering and the further mutual addition (to get the $\sin^2(\omega t) + \cos^2(\omega t) = 1$ identity), permit derivation for the canonical ellipse equation in the mentioned phase plane:

$$\frac{1}{b^2}\left(\Delta x - \frac{Ct}{B(B-t)}\right)^2 + \frac{1}{(b\omega)^2}\left(\dot{x} - \frac{C}{(B-t)^2}\right)^2 = 1, \quad (6)$$

where the centre depends on *t*, and semi-axes are determined by the amplitude-frequency response (AFR), i.e. $\omega$ and *b*. Remembering about (3), we can conclude about a shifting of the ellipse (6) together with its centre along the $\dot{x}$ axis to infinity, in the first quadrant by growth in *t* that approves instability of the phase trajectory. Also, unlimited and non-uniform changes of AFR together with other factors acting in time (such as critical heating rate or load, phase transitions, etc.) can be related to the system stability. Then, together with (4), additional expressions (like a phase equilibrium equation, and so on) can be used in a mathematical set (i.e. a system). Some examples of initial conditions and pre-determining factors are described in references [28,38,39]. Particularly, such elliptical phase trajectories was observed in [38] with a suggestion about the presence of an unknown master non-linear differential equation. Changing of AFR and the possible resonance can be investigated either approximately [19,21] or precisely. Since some solutions of (3) are known in the explicit form, direct harmonic Fourier analysis with $a(n;\omega)$ amplitude derivation is reasonable. To make this, Eq. (3) should be expressed as a trigonometric Fourier series (over the $0<t<B$ real interval) with an $a(n;\omega)$ amplitude searching to derive AFR. The case when $a(0;\omega)$ corresponds to stable oscillations (the initial ellipse), and the amplitude is determined as:

$$a_0 = a(0;\omega) \sim -\frac{2b(\cos(\omega B) - 1)}{\omega B}. \quad (7)$$

From the numerical analysis of (7), it follows that zero stable harmonics do not contain resonance elements, but the growth by *n* leads to instability (i.e. $a(n;\omega) \sim -\frac{\sin(\omega B)}{\omega^2 - M_1^2} - \frac{\omega \cos(\omega B)}{\omega^2 - M_2^2}$, at constant $M_1$ and $M_2$, coincides qualitatively with AFR from [19,43] and correlates as an unsmoothed signal with its averaged one). Comparing results from [39] with the discussed model (3) and literature data [19,43], we can conclude resonance behaviour in practice with beats in a «soft» system (the resonance frequency $\omega=2\pi n/B$, where $n$ is a natural number). If we also take account of the above-mentioned transition from (3) to $x_0(1-\exp(-qt))$ at a small $t$ or isothermal conditions (together with $b\cos(\omega t) \to b$), the «viscoelastic» Kelvin-Voigt function can be expanded into the Fourier series. Then, determining $a(t;\omega)$ AFR over a variable time interval as a two-parameter function enables a qualitative description of the reversible resonance maximum, achieving at an acting frequency. It is equivalent to the description of frequency-driving $\beta$-relaxation in amorphous materials at DMA [24,25], and all the terms in a Fourier series become meaningful from the applied standpoint as a response from linked oscillation pendulums (or molecular oscillators, deforming areas) with typical interference (collective addition or subtraction of coherent waves) and resonance [28]. The next, non-isothermal creep of different materials using the mentioned expressions from this paragraph will be analysed below.

*3.2. Non-isothermal creep at different heating rates and static loads*

In non-isothermal creep tests on AMAG-183 metallic glasses, the deformation scenario (1) appears with the linear correlation coefficient exceeding 0.9 (and $C=400$ mm·s, $B=141$ s, $V_T=3$ K/s, gravity force (i.e. load) $mg=17$ N, $x_o=0.1$ mm). Due to the loading specific, harmonic oscillations, whose amplitude may be neglected, can appear on some experimental curves, or their analysis is possible with (3). Similarly, different personal values of $C$ and $B$ have been determined at various heating rates and loads with the same correlation order. Herewith, an upward-convex bend (irreversible accelerated deformation) is repeatedly observed in $x(t)$ experimental plots, mainly at 17 N loading with 0.5 or 2 K/s strictly constant heating rates (Fig. 2a), and that can be caused by phase-structural instability, demanding analysis in frames of the proposed model (see section 3.1). Other loads and heating rates do not provide such effect, but $B$ lifespan increases or becomes shorter (see Fig.2b) with variations of the maximum possible elongation and *C* parameter.

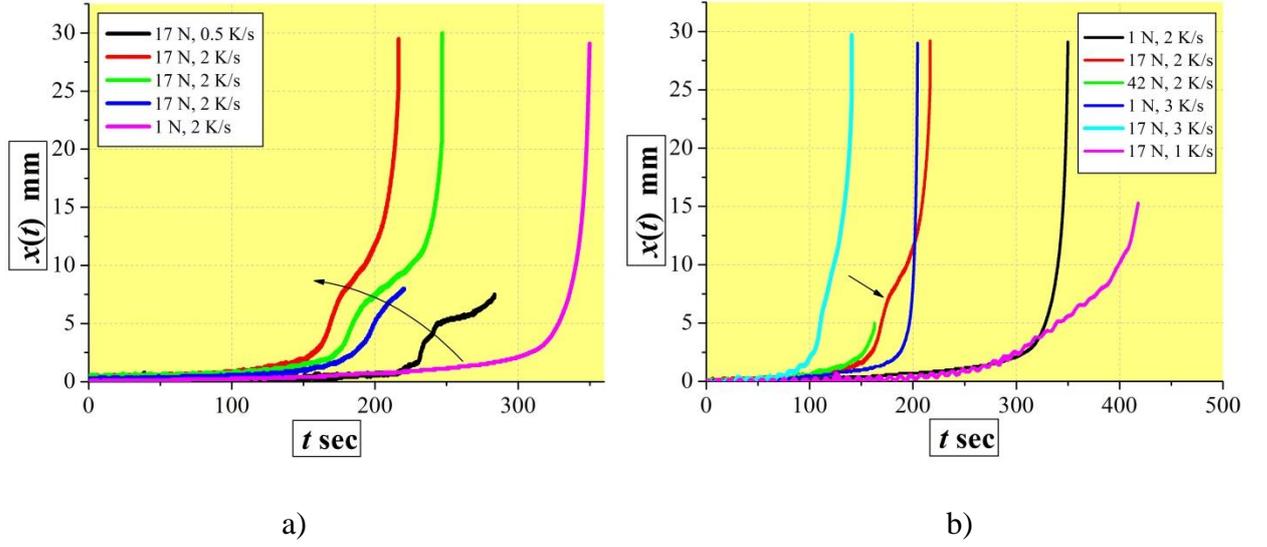

a)          b)

Fig.2. Deformation curves *x*(*t*) of AMAG-183 MG, recorded under different heating rates and loads: a) bending (the trend of instability is mentioned with the arrow). Blue and black curves correspond to specially unfinished tests to show independency between bending and boundary conditions; b) a set of curves plotted by different experiments (the arrow notes a bending segment)

Similar experimental behaviour was described in [44] on the nickel alloy with a proposed hypothesis about sum of deformation terms, but there were no suggestions on a master equation (like Duffing one) and its accompanying analysis.

To determine analytical relationships between $V_T = \Delta T / t$ heating rates, $\Delta F$ loads, and the plot bending, we use Eq. (11) rewritten for the $C_{DMA}$ thermal capacity from [28] at the statically acting force condition ($\omega = 0$), i.e.:

$$C_{DMA} = C_{TMA} = -T\left(\frac{\partial^2 G}{\partial T^2}\right)_F = -\frac{CT}{BV_T(BV_T - \Delta T)}\left(\frac{6mC}{\left(B - \frac{\Delta T}{V_T}\right)^4} + A\omega\cos\left(\omega\frac{\Delta T}{V_T}\right)\right), \quad (8)$$

where $G$ is the Gibbs energy, depending on $\Delta F$ and $T$ variables by a process (1). Using an expression for the reaction force (see Eq. (4) in [28]) as $\Delta F = F_{load} - F_{react.} = \dfrac{2mC}{\left(B - \dfrac{\Delta T}{V_T}\right)^3}$ (9), we provide the second derivative of $G$ in (8) as $\left(\dfrac{\partial^2 G}{\partial T^2}\right)_F = \dfrac{3C\Delta F}{B(BV_T - \Delta T)^2}$ (10). Since we analyse an arbitrary variation of the reaction force at non-isothermal creep, the $G$ function and its derivatives are considered on the entire process sub-plane rather than along a single $F(T)$ path. And, as for Eq. (10), it should be integrated twice to obtain the $G$ primitive (i.e. inverse derivative) function. In result, the Gibbs energy for the bending process $x(t)$ will be expressed as:

$$G = \frac{-3C\Delta F \Delta T}{B^2 V_T} - \frac{3C\Delta F}{B}\ln\left(\frac{BV_T - \Delta T}{BV_T}\right), \qquad (11)$$

and the equality between its $\left(\dfrac{\partial^2 G}{\partial F \partial T}\right)_T$ and $\left(\dfrac{\partial^2 G}{\partial T \partial F}\right)_F$ mixed derivatives confirms the validity of performed operations together with the whole approach (i.e. $G$ is the differential form on the $F$-$T$ plane). With differentiation (11) by $F$, it is possible to describe the bending of (1) in the expected phase transition area for the amorphous alloy as Eq. (8) follows from the Clausius-Clapeyron relation [28]:

$$\Delta l(T) = \left(\frac{\partial G}{\partial F}\right)_T = \frac{-3C\Delta T}{B^2 V_T} - \frac{3C}{B}\ln\left(\frac{BV_T - \Delta T}{BV_T}\right). \qquad (12)$$

Joint numerical analysis of (12) and the bending segment in $x(t(T))$ demonstrates their identity (accelerated growth) at higher temperatures, and different $V_T$ affects compatibility together with the $\Delta l$ curvature. The maximal achieved correlation coefficient between both functions is 0.9, but there are possible curves with lower values. Among the $V_T$ numbers, one can detect those, providing two-point crossing (double roots) between (12) and (1) with the mutual, experimentally approved, curve switching. It is known [45] that a functional jump of the first derivative by $G$ (i.e. (12) scenario) relates to the first order phase transition in a system.

Herewith, heat emission can appear that is described with (8) for calorimetry (similar to [28]), and the entire process corresponds to crystallisation of an amorphous structure under certain loads and heating rates. Additionally, comparison of the first and second derivatives for (1) and (12), together with corresponding reaction forces (derived from Newton's laws [39]), demonstrates irreversible accelerated deformation and material softening by (12) scenario. It is known [45-47] that, for the first-order phase transition, calculation of pseudo-critical exponents may be carried out by its relationship with thermodynamic potentials. As thermal capacity at $F$ constant loading force is a characteristic parameter, its behaviour during the bending can be estimated. To make this, Eq. (10), multiplied by «–T», should be analysed in a view:

$$-T\left(\frac{\partial^2 G}{\partial T^2}\right)_F = -\frac{3CT\Delta F}{B(BV_T - \Delta T)^2} = C_F \sim -\frac{BV_T}{(\Delta T - BV_T)^2} - \frac{1}{\Delta T - BV_T}, \quad (13)$$

considering that two powers of $(\Delta T - BV_T)$, i.e. $\alpha_1=2$ and $\alpha_2=1$ are critical exponents, respectively. One can see from (13) that the calculated temperature interval overlaps the Curie and crystallisation points of the investigated alloy [48], and critical exponents differ from the pure isothermal paramagnetic transition (when $C_F \sim 1$) [46] due to the thermal impact both on mechanical and magnetic properties. Behaviour of (13) by the absolute value is equivalent to Curie-Weiss law (especially if we neglect $\alpha_1=2$ power in smallness order at $\Delta T \to BV_T$), and relation between magnetocaloric and thermal systems is known referring to critical exponents theory (where $C_H$ ferromagnetic thermal capacity under the $H$ field is proportional to $C_F$, i.e. $C_H \sim C_F$) [46]. Since equation (12) also involves the difference between $BV_T$ and $\Delta T$, the bending intensity depends on the optimal (limited) value of $V_T$ within the (0; +∞) range, along with material-specific properties, described by $C$ and $B$. It is experimentally approved with data in Fig.2b at 17 N of load and 1, 2, or 3 K/s heating rate. As both magnetic and thermomechanic properties conjugate in the AMAG-183 MG under mentioned conditions, then we have to additionally approve the mutual impact of spin and interatomic reorganisations on the bending effect. Note that outside the phase transition region, equations (8), (10) – (13) are not applicable, hence we will consider the bending together with its modelling in this manner.

Experimental verification can be started with a comparative study of deformation curves for AMAG-183 amorphous and Al-Fe crystalline ribbons under the same conditions (for example, at $C$=144 mm·s, $B$=306 s, $V_T$=2 K/s, $mg$=1.5 N the gravity force depending on the tensile stress, $x_o$=0.1 mm, 0.9 as the correlation coefficient). In the Fig.3, non-isothermal creep curves of the aluminium polycrystalline alloy are depicted.

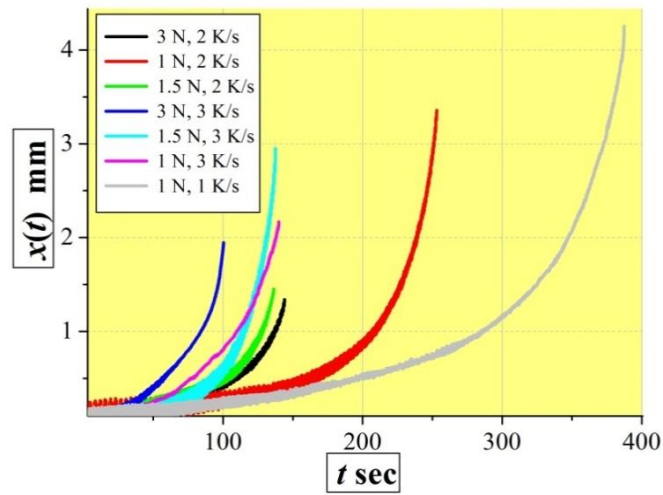

Fig.3. Deformation $x(t)$ relationships of Al-Fe polycrystalline alloy at different heating rates and statically acting loads

As shown in Fig.3, crystalline specimens demonstrate a relatively less plasticity compared to metallic glasses, although upper strain limit shorts similarly by the iterative growth of heating rate and loading. Herewith, the bending effect typical to MG does not appear in the curves that can be related to the absence of phase transitions (like magnetic or atomic ones) in this material due to its primary crystallisation (Fig.4).

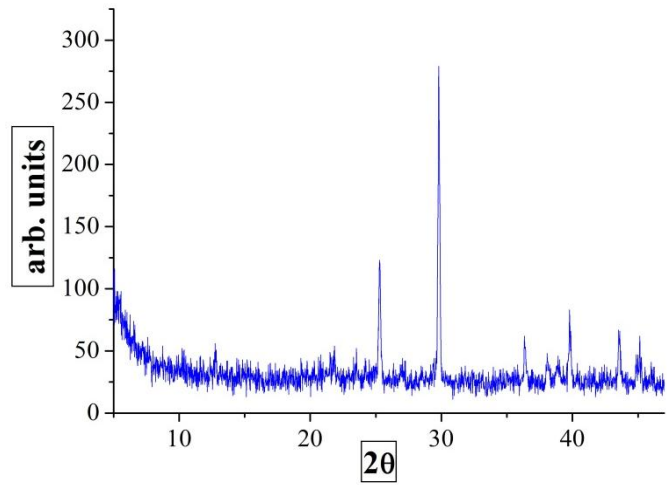

Fig.4. XRD response from the Al-Fe polycrystalline ribbon alloy

Further investigation into the relationship between initial experimental conditions and the bending of creep curves was performed on AMAG-183 specimens, previously annealed at 673 K for 10, 25, and 38 minutes, with 788 K being identified as the crystalline temperature [48]. In the Fig.5, deformation creep curves, obtained after the different annealing stages, are plotted.

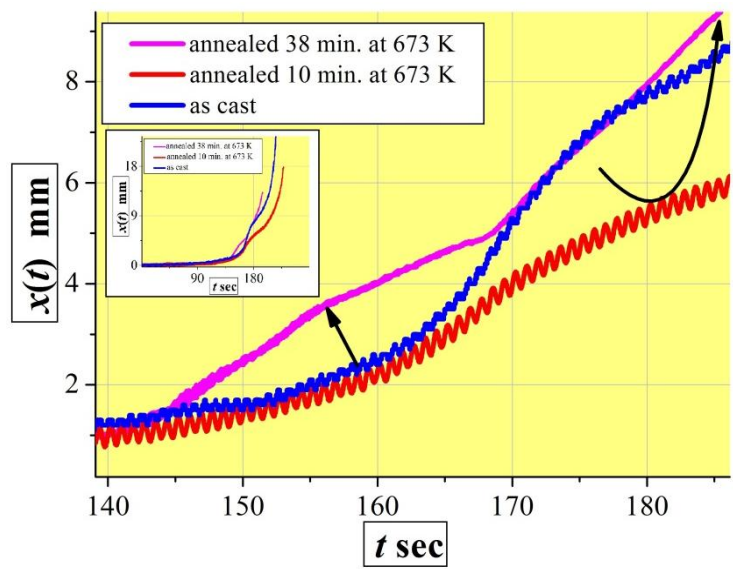

Fig.5. Creep curves of as-cast and annealed AMAG-183 specimens. Arrows indicate directions to plot smoothing, and the whole experimental interval is shown with an insert.

Fig.5 testifies about 1.5 times curve smoothing after 10 min. annealing at 673 K with the further 1.5-2 times extra aligning since 38 min. of the same temperature treatment regime. Expanding the specimen set for use in 10-38 min. annealing times (at 673 K) provides the plot distribution between the mentioned curves (not all of them are depicted to avoid the pointless overlapping). This behaviour of experimental graphs testifies to a system shift between two positions that can be described with (1) and (3) parameterization in energy and work coordinates [40]. Annealing and creep tests induce certain structural reorganisations within the material. A treatment of as-cast AMAG-183 ribbons (Fig.6a) at 673 K for 10 minutes does not lead to structural crystallisation (Fig. 6b) even after the whole creep test, and only with the pre-annealing in 38 minutes (at the same temperature), using a further non-isothermal creep experiment, diffraction peaks appear (Fig. 6c,d).

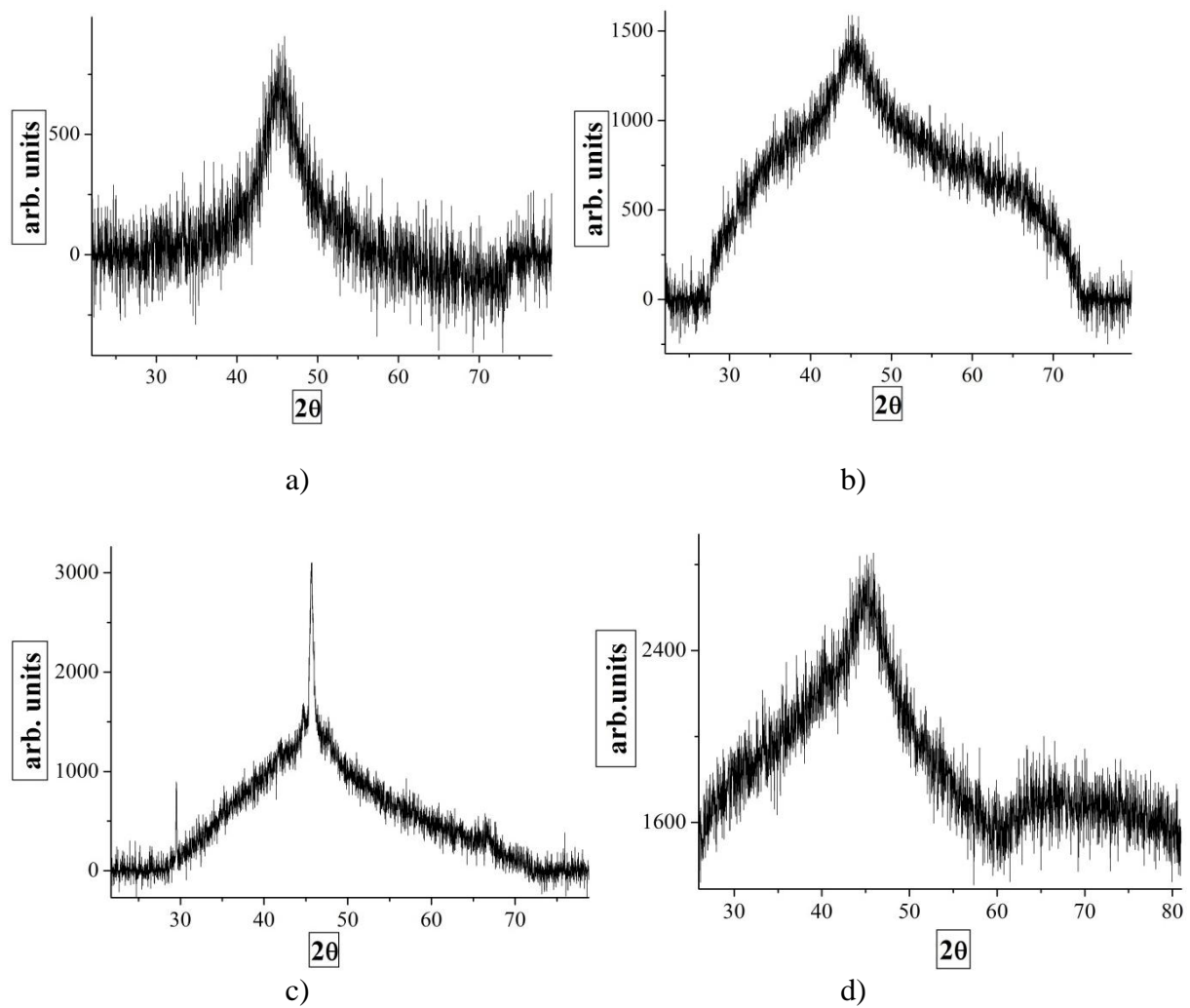

a)            b)

c)            d)

Fig.6. XRD data for AMAG-183 alloy: a) as-cast specimen; b) after a non-isothermal creep test (up to the fracture) of a pre-annealed (in 10 min. at 673 K) ribbon; c) the same as in b), but the treatment was carried out in 38 min. at 673 K; d) annealed sample in 38 min. at 673 K before the creep test

Also, there is a relation between non-isothermal creep or annealing and the wideness of ferromagnetic hysteresis loops, changing due to anisotropy and structural reorganisations [48]. Thus, data comparison permits to make a conclusion about the curve bending as a result of interatomic and magnetic reorganisations (partial relaxation, phase transitions), arising due to quite intensive thermo-mechanical impact over time. Temporal annealing under pre-crystallisation temperatures contributes to more uniform structural relaxation that is observed as the smoothing of curve bends. If we note the Fourier series expansion for (1) by normal frequencies (see 3.1 section), bending element (12) can be interpreted as a resonance of magnetic (spin-orbital) and atomic subsystems in the alloy under external mechanical impact and heating (i.e. infrared photonic flux absorption with the further phonon excitation spectra). Bending magnitude before and after annealing depends on *C* and *B* structural parameters (at determined $V_T$ with $\Delta T$), and violation of mathematical or physical stability is provided by their changing under the *F* acting load. As noted above, this reasoning (from the 3.2 section) is valid only in frames of the Clausius-Clapeyron relation [28], i.e. at phase transition, observing with applied methods (diffractometry, calorimetry, magnetometry, etc.). Beyond the phase transition and shift of equilibrium, equations and solutions from the 3.1 section are still useful for other materials in their creep analysis. Non-isothermal creep testing can also be applied as a method to determine phase transition temperatures in MG and other amorphous materials under higher heating rates and loads, which are not accessible with techniques like DMA or TMA.

*3.3. Non-isothermal creep with a further polymorphic transition in the fractured material structure*

At non-isothermal creep testing of the AMAG-200 ribbon alloy, deformation curves are precisely described with (1) (i.e. 0.98 correlation coefficient, $C=670$ mm·s, $B=650$ s, $V_T=1$ K/s, $mg=1$ N load, and $x_o=0.1$ mm) like for the AMAG-183 one. Herewith, after the fracture of a ribbon specimen, its upper free part melted with the further rolling up (opposite to the gravitational field) and forming the drop with orange light radiation (~$10^3$ K by the colour temperature scale), and that ended by mass solidification without visible glow (see Fig.7).

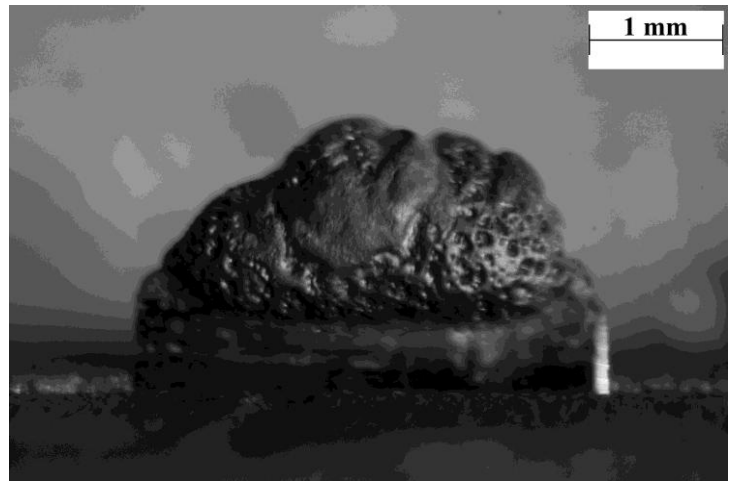

Fig.7. The solidified metal drop at the upper free part of the fractured AMAG-200 ribbon specimen

This process systematically occurs in the first minute after the end of a creep test both inside the furnace chamber (with passive temperature growth) and its outside (opened pot already without heating), i.e. stored heat, received after the fracture of a specimen, spends to the further phase transition in its upper free part. XRD testifies (Fig.8) that on the drop surface and its depth there are no molecular bonds which undergo the liquid state transition at $10^3$ K (mainly, crystalline oxides of the base atomic components with $1.5 \cdot 10^3$ K or higher melting point take place in the spectre).

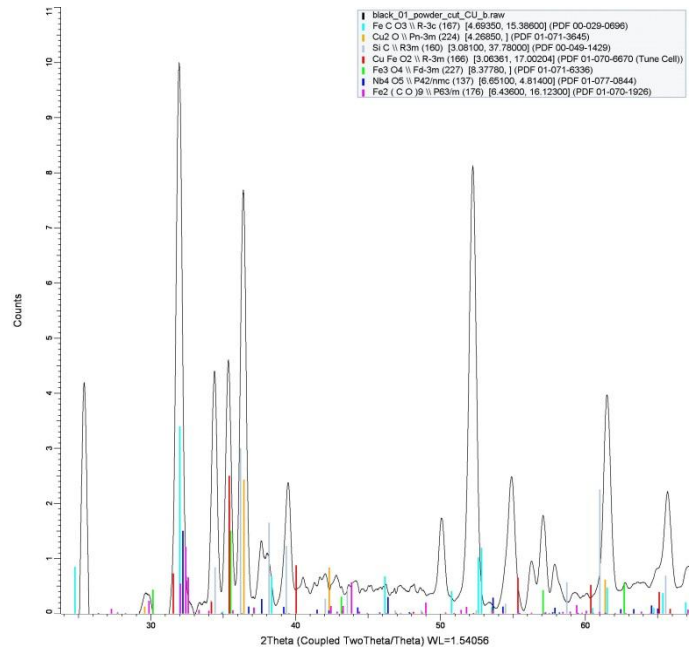

Fig.8. An XRD pattern of the solidified drop at AMAG-200 specimen matrix

Possible explanation for the observed effect is the presence of more complicated (triple or multiple) crystalline phases with the lower liquidus point and non-linear stoichiometry, based on the main alloy components [49]. The same process is not observed in the AMAG-183 system, and that indicates specific properties of the AMAG-200 alloy outside the range, wherein the Duffing equation can be applied.

*3.4. Non-isothermal creep of amorphous micro-rods with a glass cover*

Non-isothermal creep of amorphous rods covered by glass is described with (1) at 0.95 correlation coefficient, $C$=250 mm·s, $B$=960 s, $V_T$=1 K/s, $mg$=0.1 N, $x_o$=0.1 mm. Herewith, deformation is completely smooth (i.e. no bends in $x(t)$ curves) despite the combined visco-brittle fracture of a rod with its glass coating, melting, and cracking (see Fig.9), forming lateral tension zones near the contact boundary between glass and alloy [50].

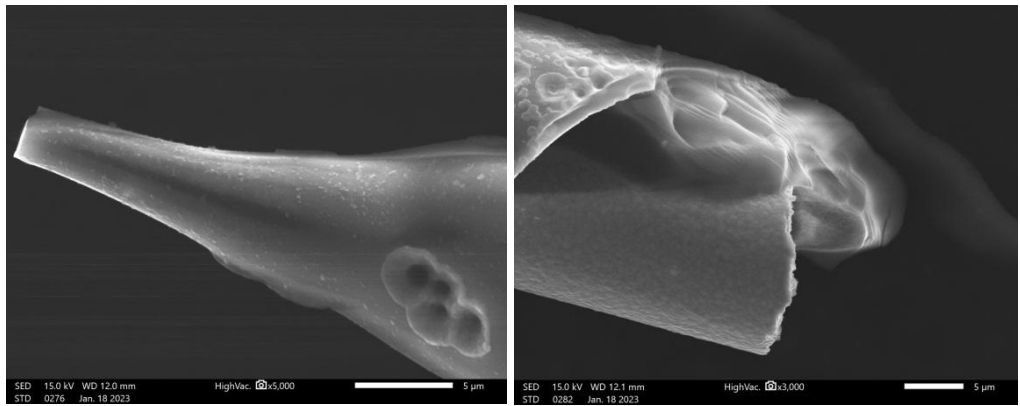

a) b)

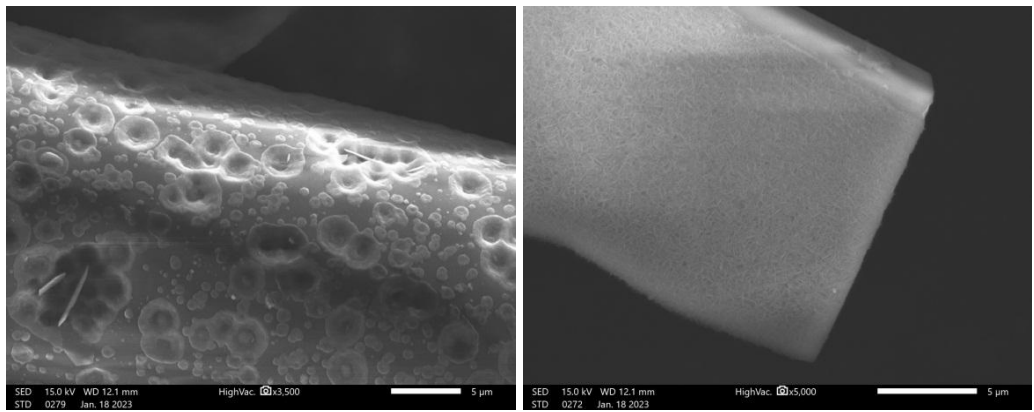

c) d)

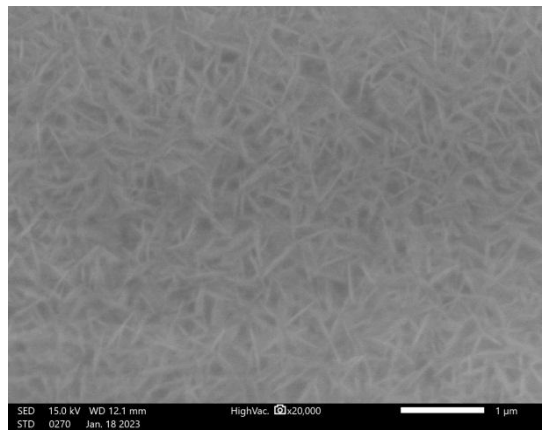

e)

Fig.9. SEM data, collected in a fracture zone of amorphous rods: a) viscous tightening in the lateral projection; b) brittle cracking of the glass coating around the rod and break; c) melting area of the glass coating; d) combined rupture of glass with specimen; e) formation of microstructure in the glass coating

Similar to AMAG-200, whose deformation is described in the 3.3 section, the shift of a melting temperature point occurs in the glass coating. In particular, in Fig. 9a,c, glass melting zones are visible despite the heating temperature not exceeding $10^3$ K during the whole experiment (and the same phase parameter is obviously higher for any $SiO_2$ systems regardless of defects [51]). This testifies to the impact of interatomic bonding in the interphase boundary between glass and alloy on local thermodynamic processes.

*3.5. Non-isothermal creep of silk fibres*

Non-isothermal creep tests of silk fibres are also modelled using (1) at 0.98 correlation coefficient ($C$=24 mm·s, $B$=442 s, $mg$=0.1 N, $x_o$=0.1 mm, $V_T$=1 K/s), but at 2 and 3 K/s heating rates a $-Dt$=-0.0182$t$ linear term should be included in the model, related to the Duffing equation, due to the experimental linear decrease. Also as AMAG-200 alloy, silk fibre structures undergo the fragment melting after the fracture of specimens (Fig. 10a), and partial weight loss is possible (Fig. 10b), but the interatomic structure stays disordered (Fig. 10c).

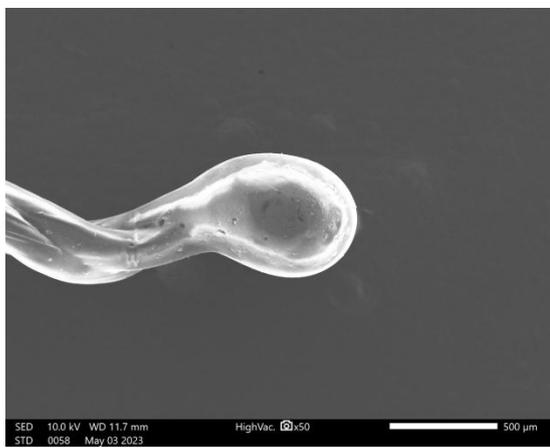
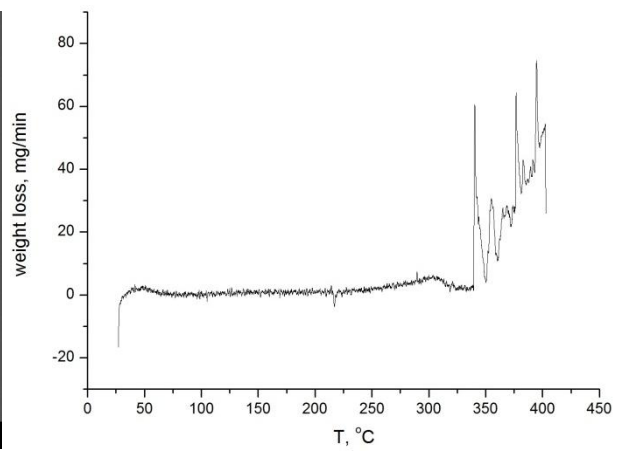

a)                      b)

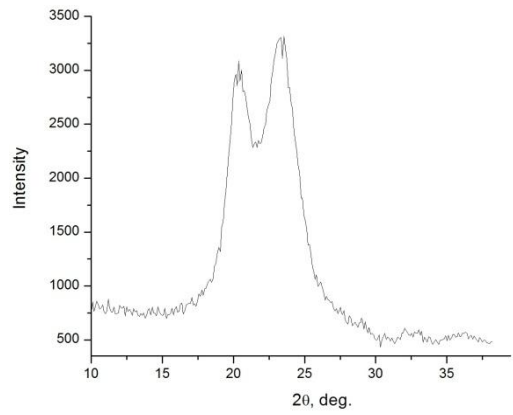

c)

Fig.10. Behaviour and condition of an investigated silk fibre: a) fragment melting and drop formation; b) a thermogram without mechanical load; c) XRD data

Mentioned deformation behaviour can be explained in this manner: with temperature growth from the room values, fibres tighten up and adapt to the external load, elastically resisting like engineering pulleys, and that is described with the negative linear term in (5a). However, the further combined action of higher temperatures and constant load leads to plastic structural deformation, which is modelled with a rational function in (1) or (3). Due to reorganisation and breakdown of molecular bonds, plastic deformation dominates over elastic one, and that provides the asymptotic tendency for $x(t)$ becoming to infinity.

## 4. Conclusion

Deformation scenarios of non-isothermal creep (including DMA and TMA) can be described with solutions proposed here for the master Duffing equation. In certain cases, specific deviations (such as bending or melting), caused by phase transitions in studied materials, are observed. As the presence of (1) or (3) is determined by a force distribution in the systems, application is universal for these analytical relationships together with the Duffing master equation, and extended models can be supplied with state equations (like Clausius-Clapeyron and others). Using the proposed model, a relationship between thermal and magnetic phase transitions can be established through the critical exponents approach. Founded non-linear solutions agree with viscoelasticity theory and enable transitions to Maxwell or Kelvin-Voigt relationships. A transition point between elastic and plastic states with the further fracture is expressed using both physical and phase portraits (i.e. with Eq.6) methods that determine stability of the whole system. Wave and resonance qualitative interpretations for thermomechanical effects [28,38,39] (such as $\beta$-relaxation) are analytically specified in this work using the Fourier series transformation of (1) and (3).

**Funding:** No funding was received to assist with the preparation of this manuscript.

**Declaration of competing interest**

The authors have no relevant financial or non-financial interests to disclose.

**Acknowledgements**

The results were partially obtained at the Center for Collective Use of Scientific Equipment at Derzhavin Tambov State University. Magnetometry was performed at Magnetic Materials Research Laboratory (Lomonosov Moscow State University). Authors thank P.V. Andreev (Lobachevsky State University) for assistance with some XRD measurements.

**Appendix**

(A.1)

To prove _T.1._, the second temporal derivative of (1) with the further exchange from $t$ to $x$, using the inverse $t = \dfrac{B^2 \Delta x}{C + B \Delta x}$ [28] function relative to (1), should be calculated. Then, the derived expression for $\ddot{x}(t(x))$ is substituted into equation (2), and the fractions are algebraically simplified by reducing to a common denominator and performing mutual subtraction of opposite terms. In result, (2) becomes to a 0=0 identity.

(A.2)

To prove that $x(t) = x_0 + \dfrac{Ct}{B^2 - Bt} - b\cos(\omega t)$ is a solution in T.2, an exchange from $t$ to $x$ can be made with $t = \dfrac{B^2 \Delta x}{C + B \Delta x}$ expression, where $\Delta x = x - x_0 + b\cos(\omega t)$, and $P_3(x;t) = P_3(\Delta x)$ is a function of $\Delta x$, i.e. an extended cubic polynomial by $x - x_0 + b\cos(\omega t)$ should be considered in the left part of the Duffing equation.